\documentstyle[11pt,newpasp,twoside,epsf]{article}
\reversemarginpar
\input{tcilatex}

\begin{document}

\title{The Relationship Between Masers and Massive Star Formation: What Can
Be Learned from the Infrared? }
\author{James M. De Buizer}
\affil{Cerro Tololo Inter-American Observatory, National Optical
Observatory, Casilla 603, La Serena, Chile.}

\begin{abstract}
The infrared represents an alternative wavelength regime in which to study
the environments of maser emission, while at the same time complementing the
information obtained through radio techniques. The near infrared (1-2 $%
\micron$) yields information on outflows, shocks, and reflected dust
emission, while the thermal infrared (3-30 $\micron$) yields information on
the thermal dust distribution around stars. Thus, the infrared regime yields
important clues in determining whether masers exist in shocks, outflows,
circumstellar accretion disks, or in the dense medium close to protostars.
\end{abstract}

\section{How Does Infrared Complement Radio?}

Observations of centimeter radio continuum sources in the vicinity of
galactic masers established the idea that masers are related to young
massive stars. However, massive stars of spectral types later than B3 do not
generate enough ionizing flux to create observable UCHII regions, given the
sensitivities of modern centimeter radio telescopes and the kiloparsec
distances to typical massive star forming regions. Furthermore, accurate
astrometry from connected element interferometers has shown that masers are
often not coincident with UCHII regions. Therefore, in order to figure out
exactly how masers are related to young massive stars, and to observe the
environments of the masers, one needs to observe at wavelengths other than
centimeter. However, there is generally large extinction due to dust and gas
in the environments near massive stars that are in the process of forming.
This makes observing young massive stars difficult or impossible at
wavelengths less the 1 $\micron$. One could use far-infrared and
sub-millimeter instruments, but these technologies presently do not have the
spatial resolution to give detailed information on the maser environments
close to young stellar sources. This means that the near and thermal
infrared are presently the best alternative ground-based spectral regimes in
which to study the formation of the individual massive stars and their
nearby environments, and how they relate to maser emission.

In star forming regions, near infrared (1-2 $\micron$) photons are usually
photospheric emission or reflected and scattered photospheric emission off
of dust near a star. The near-infrared regime is valuable because
observations can peer through the extinction near newly forming stars, if
the stars are not too embedded. The thermal infrared (3-30 $\micron$) refers
to the spectral regime which one looks at heat wavelength information. When
referring to star formation regions, the thermal infrared traces heat
radiation emitted from circumstellar and near-stellar dust. This spectral
regime traces dust temperatures between 1000 K at 3 $\micron$ (close to the
temperature of dust sublimation) to the relatively cool temperature of 100 K
at 30 $\micron$. The thermal infrared is even less affected by extinction
than the near-infrared, making it a powerful probe of star formation
regions. Therefore, using the infrared one can study the dust distribution
and environment near a young massive star over a rather broad range -- from
the photosphere and hottest material close to a star, out to the coolest
material distributed far from a stellar source.

\section{Maser Locations and Infrared Observations}

\subsection{Embedded Sources}

One scenario for the location of masers is in the near stellar environments
of deeply embedded protostars. The work of Cesaroni et al. (1994) showed
that there was a coincidence between clumps of ammonia emission and water
masers that are offset from UCHII regions. These water masers may be marking
the locations of ``hot molecular cores'' which may contain deeply embedded
protostars that are so young that they do not have observable UCHII regions.
Cesaroni et al. (1994) showed that these sources should have temperatures
between 50 and 165 K. If these cores have temperatures greater than 100 K,
they can be seen and studied in the thermal infrared. Figure 1a shows the
direct detection of a hot molecular core at thermal infrared wavelengths
from Gemini North (De Buizer et al., in print).

\begin{figure}[t]
\plotone{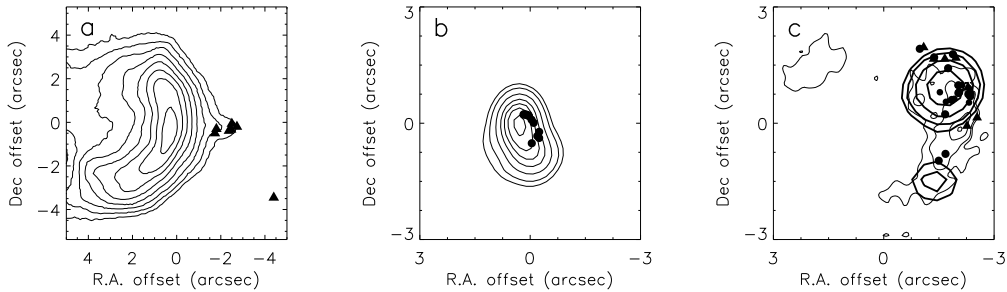}
\caption{(a) A contour map of the
cometary UCHII region G29.96-0.02 in 18 $\micron$ emission. We detect a
thermal infrared source at the location of the water masers (triangles,
Hofner and Churchwell 1996) which are offset from the UCHII\ region and
coincident with a knot of NH$_{3}$ (4,4) emission (Cesaroni et al. 1994).
(b) The 18 $\micron$ contours of G309.92+0.48, a resolved and elongated
thermal infrared source, with methanol masers overplotted as dots. The
thermal dust emission is elongated at a similar position angle as the
associated linear maser distribution. (c) Thin contours show the location of
our deconvolved 18 $\micron$ dust emission, and the thick contours show the
location of 1.25 $\micron$ (J band) emission (Persi et al. 1996) for NGC
6334 F. The J emission may be light scattered off of the dust, however this
filter contains many emission lines from shock-excited iron as well. These
observations show that there is a warm density enhancement at the location
of the masers (symbols), and perhaps the near-infrared emission is
shock-related.}
\end{figure}

More hot cores could be found by searches in the thermal infrared looking
for sources that are coincident with masers groups that are offset from, or
do not lie near, radio continuum emission. Collaborations with radio
astronomers will be needed for follow-up ammonia observations to confirm
that they are indeed hot molecular cores. Using this combined information,
models for the earliest stages of massive star formation can be constrained.
Thermal infrared fluxes alone can refine massive star formation models like
those of Osorio et al. (1999).

\subsection{Circumstellar Disks}

Another scenario for the location of masers is in circumstellar disks around
young massive stars. The work of Norris et al. (1993) originally showed that
there was a strong tendency for methanol masers to be distributed in linear
patterns, and in some cases, there are velocity gradients along these
distributions. It was determined that the simplest explanation for this
phenomenon was that the masers are located in edge-on circumstellar disks.
Likewise, in a small number of cases the spatial and velocity distributions
of water and OH masers have been best described as existing in disks
(Brebner et al. 1987; Torrelles et al. 1996; Slysh et al. 1999). If masers
do exist in disks, then they would be best seen in the infrared. De Buizer
et al. (2000) performed the first survey to detect circumstellar disk
candidates at \ thermal infrared wavelengths. An example is shown in Figure
1b, and a full discussion of these sources and their likelihood of being
actual circumstellar disks is given in De Buizer et al. (2000).

At the 1-10 kiloparsec distances of these star forming regions, high
resolution imaging is essential in the thermal infrared in order to resolve
sources that may have circumstellar disks. For those sources that are too
far away or do not have disks that are extended, indirect evidence of the
presence of disks can come from mapping outflows from these sources. If the
outflow is perpendicular to the linear distribution of maser spots, this is
good evidence that these masers exist in a disk. One way to accomplish this
is to search for signs of outflow by mapping these regions in the near
infrared with narrow band filters centered on spectral outflow indicators,
such as the 2.12 $\micron$ H$_{2}$ and 1.64 $\micron$ [FeII] spectral lines
(see Davis \& Eisloffel 1995).

\subsection{Edges of UCHII Regions}

A further scenario for the location of masers is in the dense regions at the
edges of UCHII regions. Density enhancements exist between shock and
ionization fronts of expanding UCHII regions. Material can also be swept up
to create a ``bow-shock'' at the head of a moving cometary UCHII region. A
star sitting on the edge of a density gradient in a molecular cloud will
create a cometary UCHII region as well, and the material near the head will
be warm and dense. All of these situations may give rise to masers at the
edges of UCHII regions. Some observations have supported this idea (Ho et
al. 1983; Baart \& Cohen 1985; Gaume \& Mutel 1987), and the chemistry
behind shock fronts was found to be a good location for maser emission
(Elitzur \& de Jong 1978). If the swept up or compressed material in these
regions is warm enough, it can be observed in the infrared. Figure 1c is an
18 $\micron$ Keck image that shows water, OH and methanol maser excitation
on a dusty UCHII region edge (De Buizer et al., in prep).

Further ground-based, high-resolution, infrared observations will be capable
of studying these environments of maser emission. Observations with
near-infrared spectroscopy can confirm the shocked nature of these regions
by searching for the shock-excited [FeII] lines available at 1.25 $\micron$
and 1.65 $\micron$ as diagnostics.

\section{Conclusions}

The infrared regime is a valuable complementary tool to radio and millimeter
observations. Infrared observations aid in the interpretation of activity in
star formation regions containing maser emission. The near-infrared allows
one to see photospheric emission from an exciting young massive star or to
see this emission reflected off dust in the near stellar environment.
Furthermore, there are spectral lines that allow one to perform useful
spectroscopy and map outflows and shocks associated with maser emission. The
thermal infrared allows one to probe through significant extinction in these
regions and observe the hotter circumstellar material close to the exciting
stellar sources. It also is a good probe of the warm, dense parts of UCHII
regions. Presently, 8 to 10-m class telescopes and infrared instruments
represent the best high-resolution alternative to the radio for studying the
detailed environments of maser emission located in massive star forming
regions.

\end{document}